\documentclass{phbauth}        
\usepackage{graphicx}

\begin{document}

\begin{frontmatter}

\title{
Low-Energy Spin Dynamics of CuO Chains in YBa$_2$Cu$_3$O$_{6+x}$}

\author[address1]{D.N.\ Aristov \thanksref{thank1}},
\author[address2]{D.R.\ Grempel \thanksref{thank2}}

\address[address1]{Petersburg Nuclear Physics Institute, Gatchina 188350
St.Petersburg, Russia }
\address[address2]{Service de Physique de l'\'Etat Condens\'e, CEA-Saclay, F-91191 Gif-sur-Yvette, France
}

\thanks[thank1]{Corresponding author. E-mail: aristov@thd.pnpi.spb.ru}
\thanks[thank2]{On leave from
Service de Physique Statistique, Magn\'{e}tisme et
Supraconductivit\'{e}, CEA-Grenoble, F-38054 Grenoble, France.}
\begin{abstract}
We study the  spin fluctuation dynamics of Cu-O chains in the oxygen
deficient planes of YBa$_2$Cu$_3$O$_{6+x}$. The  chains are described
by a model including antiferromagnetic interactions between  the spins
and Kondo-like scattering of the oxygen holes by the copper spins.
There are incommensurate spin fluctuations along the direction of the
chains.  The dynamic structure factor of this system is qualitatively
different from that of a quasi one-dimensional localized
antiferromagnet due to the presence of itinerant holes.  We compute the
dynamic structure factor that could be measured in neutron scattering 
experiments.
\end{abstract}


\end{frontmatter}

The oxygen deficient CuO$_x$ planes play an important
role in the physics of the high-$T_c$ compounds
YBa$_2$Cu$_3$O$_{6+x}$. These planes are composed of
Cu$_{n+1}$O$_n$ chain fragments in which Cu and O atoms form a linear
alternating chain. The physical
properties of the chain planes are dominated by the longest
chain fragments.
To a first approximation, these chains may be viewed as
decoupled from each other and from the CuO$_2$ planes. Therefore,
important one-dimensional fluctuation effects are expected in the
charge and spin excitation spectra of the CuO$_x$ planes.

The  low-energy spin dynamics of this system was studied theoretically
in \cite{ArGr} within a strong-coupling model in which every Cu site is
occupied by a hole while the average filling of the O sites $c<1$.
Holes can hop between the O sites obeying a constraint of no double
 occupancy. There is an exchange interaction between the spins of all
pairs of adjacent holes. In this model,  a spin {\it sequence} on the
 chain is not affected by the oxygen-hole motion, {\it i.e.}, there is
spin-charge separation. Due to this fact, the energy spectrum of the
system can be computed in closed form. The spin-spin correlation
function of the chains is non-trivial, however.  It is given by a
convolution of two functions.  The first one is the dynamic structure
factor of a one-dimensional Heisenberg antiferromagnet on a
``fictitious'' lattice containing a number of sites equal to the total
number of holes in the chain. The second factor is related to the
density-density correlation function of the itinerant oxygen holes and
describes the evolution of the spin {\it  positions} as determined by
 the motion of the holes. The energy scales characterizing these two
factors  are the spin-wave velocity $v_S$ and the Fermi velocity $v_F$,
respectively.
For $c\simeq0.3$, the parameterization of \cite{ArGr} gives
$v_S = 150\,$meV and $v_F = 800\,$meV  {\it i.e.} $v_S \ll v_F$.

An apropriate tool for the experimental study of the  magnetic
fluctuations is inelastic scattering of spin polarized neutrons.
The magnetic intensity is modulated along the direction of the chains
and peaks at the incommensurate position, $Q_{{\rm max}}\!=\!\pi(1+c)$. The
spectrum of magnetic excitations at $q\!=\!Q_{\rm max}$ extends up to
the energies of order of the Fermi energy of the itinerant holes.
The absolute value of the neutron cross-secton at $Q_{{\rm max}}$ was
estimated in \cite{ArGr}. Here we compute the shape of the magnetic
signal near the maximum as predicted by our model.

Up to fundamental constants, the neutron's differential cross-section
for magnetic scattering is given by the expression :

\begin{eqnarray}
     \frac{d\sigma(q,\omega)}{d\Omega_s d\omega} &\propto&
     \sqrt{\frac{E_i - \omega} { E_i}}
     {S}(q,\omega),
     \label{crosssection}
\end{eqnarray}

\noindent
with $E_i$ the energy of the incident neutron,  $q$ the momentum
transfer in the direction parallel to the chain and $\omega$ is the
neutron's energy loss. The form of the spin-spin correlation function
${S}(q,\omega)$ was found in \cite{ArGr}.
Near $q=Q_{\rm max}$ we may keep only the principal term and linearize
the dispersion relation of both spin and density fluctuations. Up to a
coefficient we have

        \begin{eqnarray}
        S(Q_{{\rm max}}+&&q,\omega) \simeq
        \int_0^\omega d\varepsilon
        \int_{-\infty}^\infty
        \frac{dk}{\sqrt{v_F|k|} }
        \frac1{
        \left[\varepsilon^2\! -\! v_F^2k^2\right]^{1/4}
        }
        \nonumber \\ &&\times
        \left[ (\omega\!-\!\varepsilon )^2
        -v_S^2(q\!-\!k)^2\right]^{-1/2}.
        \label{finalsqw}
        \end{eqnarray}

\noindent
The shape of the function ${S}(q,\omega)$ is shown in the
upper half of the Fig.\ 1.

Consider an experimental geometry in which the momentum of the
outgoing neutron is in the direction
 perpendicular to chains and all the neutrons
are collected, irrespective to their final energy.
The measured intensity is proportional to the integral
      $
      I(q) = \int_0^{E_i} d\omega
      \frac{d\sigma(q,\omega)}{d\Omega_s d\omega}
      $
Experimentally, one typically uses polarized neutrons with incident energy
 $E_i =
25\,$meV.  We show the corresponding curve $I(q)$ in the lower half of
the Fig.\ 1.

In summary, we calculated the shape of a magnetic signal from
the chains in
YBa$_2$Cu$_3$O$_{6+x}$, visible in neutron experiments.
Albeit weak, this signal could be observed in a high flux reactor.

\begin{figure}[tbp]
\begin{center}\leavevmode
\includegraphics[width=0.8\linewidth]
{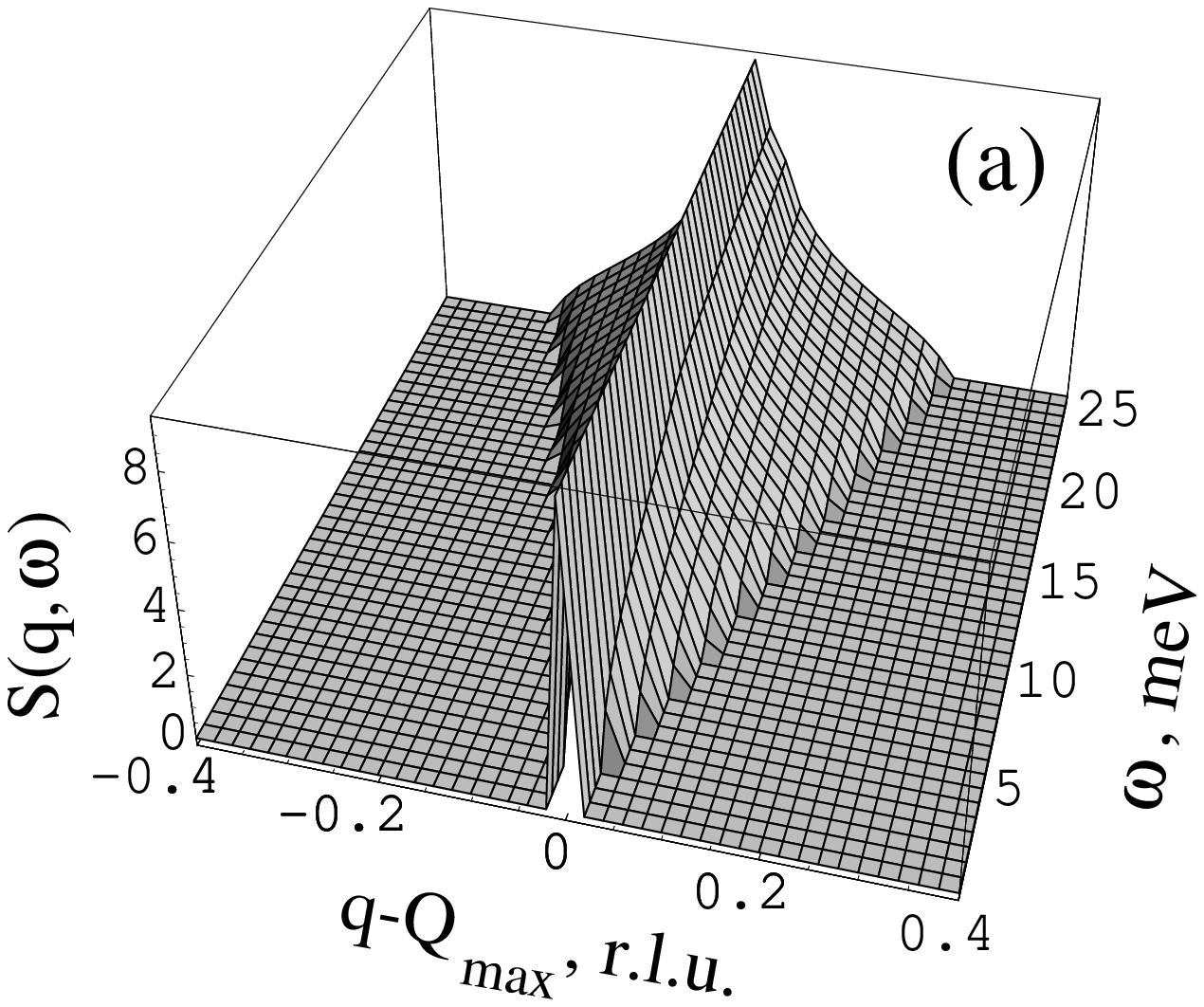}
\includegraphics[width=0.9\linewidth]
{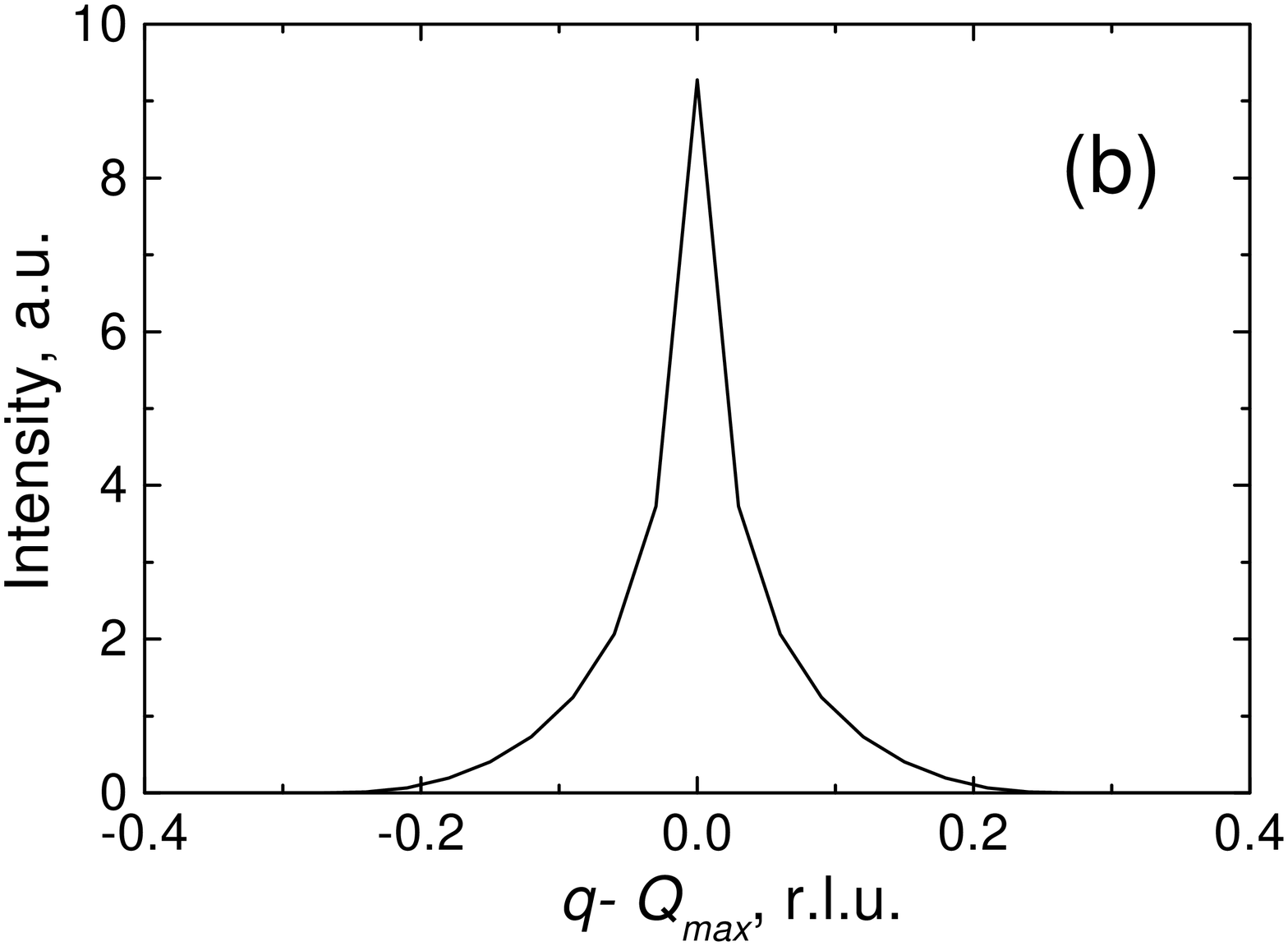}
\caption{
The dynamic structure factor of the chains $S(q,\omega)$ is shown
in the upper panel. The predicted integrated signal for an incident
neutron energy $E_i=25\,$meV is plotted in the lower part of the figure.
}\label{figurename}\end{center}\end{figure}

One of us (D.A.) acknowledges partial financial support from the
Russian State Program for Statistical Physics (Grant VIII-2) and the
Russian Program "Neutron Studies of Condensed Matter".



\begin{thebibliography}{9}
\bibitem{ArGr}
D.N. Aristov and D. R. Grempel,
  Phys. Rev. B {\bf 55}, 11358, (1997) ;
D.N.\ Aristov, {\em ibid.} 
{\bf 57}, 12825, (1998).


\end{thebibliography}
\end{document}